\documentclass[11pt,twoside]{article}
\usepackage{FUSE2004}
\usepackage{natbib}

\usepackage{epsf}
\usepackage{psfig}
\usepackage{lscape}

\begin{document}
\title{The \ion{He}{2} Lyman alpha forest: evaluation of simulated data}
\author{C. Fechner, D. Reimers}
\affil{Hamburger Sternwarte, Universit\"at Hamburg, Gojenbergsweg 112, 21029 Hamburg, Germany}

\begin{abstract}
Using an artificial \ion{H}{1} Ly$\alpha$ spectrum we simulate the 
corresponding \ion{He}{2} forest with fixed values of $\eta$ and a 
Doppler parameter consisting of a thermal and a turbulent part. 
In addition metal lines with line strengths and line density 
as expected in the case of HS~1700+6416 are superimposed. FUSE-like 
noise is added. The analysis of the simulated spectra 
in terms of Doppler profiles recovers the input $\eta = N($\ion{He}{2}$)/N($\ion{H}{1}$)$ 
with a scatter by a factor of 10. About 10\,\% 
have significantly lower $\eta$-values for various reasons. The majority of 
the extremely high $\eta$-values (up to 1000) are evidently caused 
by metal lines. We conclude that part of the scatter 
in $\eta$ of previous analyses of the \ion{He}{2} forest in 
HE~2347-4342 can be regarded as an artifact. However, we confirm 
that the correlation between small column densities (in voids) and 
high $\eta$-values is not a methodical artifact and appears to 
be a true phenomenon.
\end{abstract}

\section{Introduction}

Using the column density ratio $\eta = N($\ion{He}{2}$)/N($\ion{H}{1}$)$ the sources and fluctuations of the intergalactic ionizing continuum can be examined observing the \ion{He}{2} Ly$\alpha$ forest.
Following theoretical predictions $\eta$ is expected to be in the range of $\sim 50 - 100$ \citep{HM96, FGS98} assuming the diffuse background radiation of quasars.
Recent analyses of the resolved \ion{He}{2} Ly$\alpha$ forest towards the QSO HE~2347-4342 reveal values of $\eta$ ranging from 1 to $ >1000$ \citep{krissetal01, shulletal04, zhengetal04}.
Therefore, the authors suggest that the intergalactical UV background radiation is strongly variable on very small scales 
requiring the dominance of local sources.
However, the scatter can only be explained partly by the different spectral indices of QSOs 
\citep{telferetal02}, since the observed scales are much smaller than the typical distances between AGN.
Another finding is that absorbers in \ion{H}{1} voids show higher $\eta$-values, a phenomenon that is not yet understood.

On the basis of simulated spectra we investigate, whether the typical quality of the present data is sufficient to recover a constant $\eta$, and which effects are produced artificially by the analysis technique.
Furthermore, we examine the impact of additional metal line absorption on the results.

\section{Creating and evaluating the artificial datasets}

We generate an artificial Ly$\alpha$ forest in the redshift range $2.292 \le z \le 2.555$, which corresponds to the wavelength coverage of the lower FUSE detector segment.
The column density distribution function with $\beta = 1.5$ is adopted from \citet{kirkmantytler97}. 
The Doppler parameter distribution is described by a truncated Gaussian with $b_{0} = 27$\,km s$^{-1}$, $\sigma_{b} = 8.75$\,km s$^{-1}$, and $b_{\mathrm{min}} = 10$\,km s$^{-1}$ as observed towards HS~1700+6416 in good agreement with \citet{huetal95}.
In addition, the parameters of the simulated line sample are correlated by $b_{\mathrm{min}} = 10.5 + 1.3\cdot(\log N - 12.5)$ following \citet{misawa02}.
The resolution of $R = 40\,000$ and the signal-to-noise ratio of $S/N \sim 100$ were chosen to match the characteristic of high-resultion spectra taken with VLT/UVES or Keck/HIRES.

The \ion{He}{2} Ly$\alpha$ forest is computed from the artificial \ion{H}{1} data using $\eta = 80$, the mean value as found by \citet{krissetal01}, and a temperature of $10^{4}\,\mathrm{K}$, which is cooler than the average $2\cdot 10^{4}\,\mathrm{K}$ measured by \citet{ricottietal00}, but consistent with $b_{\mathrm{min, H\, I}} = 10$\,km s$^{-1}$.
Tests with higher temperatures reveal no difference in the results. 
Metal absorption lines were added with strength and distribution as expected in the spectrum of HS~1700+6416.
Line parameters were taken from the observed FUSE spectrum of this QSO.
The resolution of $R = 15\,000$ and the signal-to-noise ratio of $S/N \sim 5$ were chosen to match the typical values of real data.

\begin{figure}[!ht]
  \plotfiddle{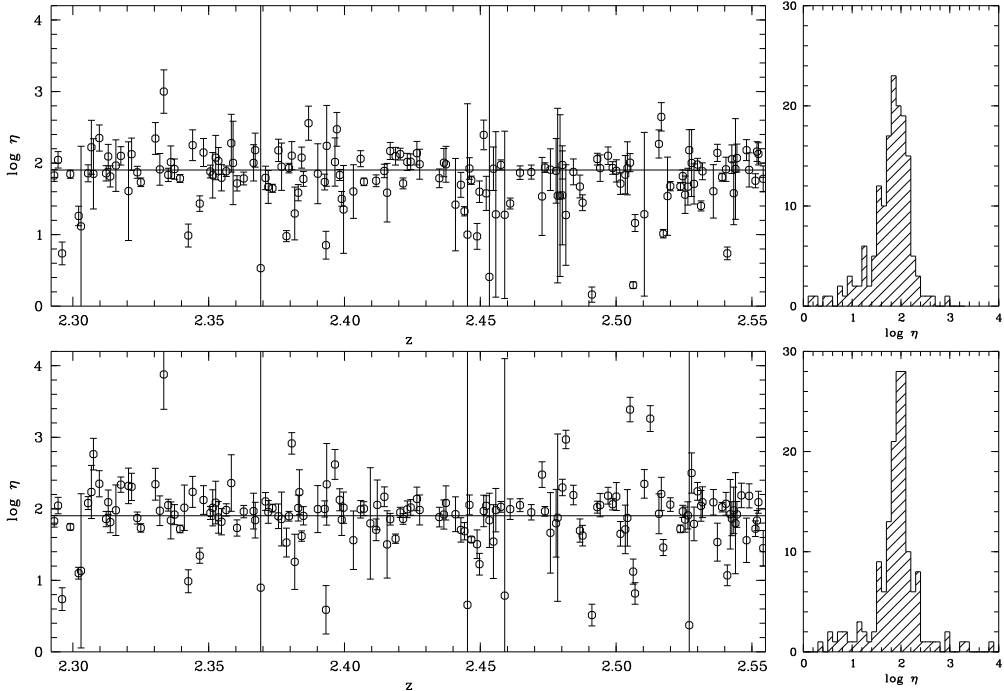}{8.5cm}{-90}{50}{50}{-205}{280}
  \caption{Column density ratio $\eta$ for the \ion{He}{2} spectrum without (upper panel) and with (lower panel) additional metal line absorption. The horizontal line indicates the expected value $\log \eta = 1.90309$ corresponding to $\eta = 80$.}
  \label{fig1}
\end{figure}

The artificial \ion{H}{1} Ly$\alpha$ forest was analyzed by fitting Doppler profiles.
The statistical properties of the sample are recovered.
The \ion{He}{2} Ly$\alpha$ forest lines were treated the same way using the derived \ion{H}{1} parameters.
We fixed the line redshift and the $b$-parameter assuming pure turbulent broadening like \citet{krissetal01}. 
The resulting $\eta$-values are shown in Fig. \ref{fig1}.

\section{Results}

From the artificial spectrum without any metal line absorption we find a mean value of $\langle\log \eta\rangle = 1.77 \pm 0.43$, which is less than the expected 1.903.
From the spectrum with additional metal lines we get $\langle\log \eta\rangle = 1.90 \pm 0.49$.
These values suggest that a scatter of about 0.5 dex in $\eta$ is due to the specifics of method.
To analyze in detail what causes the extreme deviations we refer to $\eta$-values outside the range $\log \eta = 1.903 \pm 0.500$.
Without metals 2.6\,\% of the absorbers have high and about 14\,\% low values.
The reasons for the extreme $\eta$-values are 
\begin{itemize}
  \item the \ion{He}{2} line is narrower than the adopted pure turbulent \ion{H}{1} $b$-parameter
  \item weak \ion{H}{1} lines, whose parameters are estimated incorrectly or which are misidentified
  \item blending with other \ion{He}{2} lines
  \item line saturation leading to erroneous column densities
\end{itemize}
The last three points can lead, in principle, to high as well as low $\eta$-values, while the first one produces only low $\eta$-values.
In contrast to \citet{shulletal04} who found a correlation between low density \ion{H}{1} absorbers and high column density ratios using an apparent optical depth method, we find weak \ion{H}{1} lines often leading to small $\eta$.
This can be explained by the large uncertainties of weak lines, concerning the line position.
Thus, the column densities of the \ion{He}{2} lines with fixed redshifts are underestimated.  

Generally, the presence of metal lines leads to higher $\eta$-values as can be seen from the lower panel of Fig. \ref{fig1}, where 80\,\% of the extremely high $\eta$-values are due to metals.
Furthermore, there are fewer absorbers with low $\eta$, since blending with metal lines counteracts the other effects.
Considering the total sample 5\,\% of the $\eta$-values are contaminated by metals.
About 15\,\% of the superimposed metal lines cause higher $\eta$-values.


\end{document}